\documentclass[journal=apchd5, manuscript=letter, layout=traditional]{achemso}
\setkeys{acs}{articletitle = true}

\usepackage{graphicx,amsmath, amsfonts,amssymb,subfigure,dsfont,color,ulem}

\newcommand{\vect}[1]{{\mathbf{ #1}}}

\SectionsOn
\SectionNumbersOn

\title{Plasmonic edge states: an electrostatic eigenmode description}

\author{Daniel E. G\'omez}
\email{daniel.gomez@rmit.edu.au}
\affiliation{ARC Centre of Excellence for Exciton Science, RMIT University, VIC, 3001, Australia}

\author{Yongsop Hwang}
\affiliation{Nanophotonics Research Centre, Shenzhen University, Shenzhen, 518060, China}
\alsoaffiliation{School of Engineering, RMIT University, Melbourne, Victoria, 3001, Australia}

\author{Jiao Lin}
\affiliation{School of Engineering, RMIT University, Melbourne, Victoria, 3001, Australia}

\author{Timothy J. Davis}
\author{Ann Roberts}
\affiliation{School of Physics, The University of Melbourne, VIC 3010, Australia}


\begin{document}

\begin{abstract}
We consider periodic arrangements of metal nanostructures and study the effect of periodicity on the localised surface plasmon resonance of the structures within an electrostatic eigenmode approximation.
We show that within this  limit, the collective surface plasmon resonances of the periodic structures can be expressed in terms of superpositions of the eigenmodes of uncoupled nanostructures that exhibit a standing--wave character delocalised across the entire periodic structure.
The formalism derived successfully enables the design and accounts for the observation  of plasmonic edge-states in periodic structures. 
\end{abstract}

\maketitle

\section{Introduction}

Metal nanostructures can support light--driven coherent oscillations of  electrons known as localised surface plasmon resonances. 
The frequency of these resonances is largely controlled by the geometry of the nanostructure, and by the chemical composition of both the metal  and that of its immediate surrounding.
When these nanostructures are in close proximity to each other, \textit{plasmon hybridisation} \cite{Prodan_S2003a} can take place, leading to modifications of the optical properties of the nanostructures which result from the influence  on the electron oscillations of the evanescent fields from neighbouring elements.
This plasmon hybridization has been shown to lead to many interesting phenomena such as   plasmonic analogues of Electromagnetically Induced Transparency \cite{Liu_NM2009a} and Fano resonances \cite{Fano_PR1961a,Lukyanchuk_NM2010a,Miroshnichenko_RMP2010a},
which in turn have seen applications in optical data processing \cite{Guo_NL2015a}, and sensing \cite{Lukyanchuk_NM2010a}.

When these plasmonic structures are arranged  periodically in space, interesting effects arise due to excitation of extended modes involving the  entire lattice \cite{Adato_POTNAOS2009a}.
These periodic arrangements of metal nanostructures connect plasmonics to the realm of metamaterials, which to date have been shown to lead  to novel phenomena such as negative refractive index\cite{Shelby_S2001a}, 
the prospect of optical cloaking devices\cite{Schurig_S2006a}, 
perfect absorption \cite{Ng_AN0a,Landy_PRL2008a},
mathematical operations with light \cite{Hwang_APL2016a}, 
 and 
super lensing \cite{Fang_S2005a}.
The design of optical metamaterials is often complex, requiring  numerical solutions  to Maxwell equations, or the use of genetic algorithms that target the design towards a specific application \cite{Hu_AN2016a}.

One of the simplest periodic arrangements consists of linear chains of plasmonic nanoparticles\cite{Brongersma_PRB2000a,Maier_NM2003a,Maier_APL2002a,Maier_PRB2002a}, a geometry that has been 
proposed to operate as optical waveguides \cite{Maier_NM2003a}.
For specific spatial configurations, one--dimensional particle chains can exhibit   \textit{plasmonic edge states}\cite{Poddubny_AP2014a,Sinev_N2015a,Ling_OE2015a,Wang_NJOP2016a,Downing_PRB2017a}  characterised by  significant  concentration of electromagnetic energy density in one of the ends of the chain.
These states have been studied theoretically \cite{Poddubny_AP2014a, Ling_OE2015a,Downing_PRB2017a} 
 and have been experimentally demonstrated to take place for zigzag chains of gold nanodisks using near--field microscopy \cite{Sinev_N2015a}.
Structures exhibiting these states have also been theorised to exhibit interesting non--trivial topological properties\cite{Poddubny_AP2014a, Ling_OE2015a} and can therefore serve as building blocks for topological photonics \cite{Lu_NP2014a}.

Here we present a simple nearest--neighbour approximation to describe the collective surface plasmon eigenmodes of periodic arrangements of metal nanostructures, with an emphasis on nanoparticle chains.
The theoretical model is based on a formulation of Maxwell equations in terms of optical near fields\cite{Davis_ROMP2016a}, wherein the optical response of metal nanoparticles is described in terms of self--sustained oscillations of conduction band electrons, termed surface plasmon eigenmodes.
The paper is organised as follows: 
in section \ref{sec:EEM} we introduce the general elements of the theory and in particular the implications of translation invariance in the description of surface plasmon eigenmodes.
The general results of this section are later applied to specific cases in section \ref{sec:NNA} where the near--neighbour approximation is introduced, and in section \ref{sec:4}, we illustrate how the developed formalism enables the design of structures that exhibit  plasmonic edge states, and we furthermore present an experimental realisation of these kinds of structures.
This is later followed by a discussion of the results and of the limitations of the theory.

\section{Electrostatic Approximation}
\label{sec:EEM}

The aim of this section is to  present an analysis of   systems of coupled metal nanoparticles within the electrostatic limit, that is, when the sizes of the nanostructures are much smaller than the wavelength of light. 
In this limit the coupling between nanoparticles arises from the electric near-fields of the localised surface plasmons.
In particular, we resort to the electrostatic eigenmode method (EEM)\cite{Davis_ROMP2016a,Gomez_JMCC2014a}, which enables an analytical description of the optical response of a collection of nanoparticles in terms of their mutual interactions  \cite{Davis_NL2010a,Davis_PRB2009a}.
When we consider periodic arrangements of nanoparticles, we can apply methods similar to those used in the tight--binding approximation in solid state physics\cite{Ashcroft_1976a} in which the response of electrons in an ionic lattice is represented as a linear combination of the orbitals of the individual atoms, a good basis set to represent the electronic states in a  crystal. 
In our case we use the surface charges of the localised surface plasmon resonances that play a similar role to the localised atomic orbitals, and we impose the periodic conditions on their linear combinations which is equivalent to the formation of Bloch waves. 
As we show below, it is the phase of these “Bloch waves” and the boundary condition imposed on it that determines the properties of the plasmonic lattice and leads to the edge states.

Within the electrostatic approximation\cite{Davis_NL2010a,Gomez_JMCC2014a},  the surface plasmon resonances of nanoparticles are described in terms of surface charge distributions $\sigma_{\alpha,j}(\vect r)$ (and surface dipole distributions $\tau_{\alpha,j}(\vect r)$, both forming a bi-orthogonal set, each distribution is indexed with $j$), corresponding to self--sustained oscillations of conduction band electrons in the surface points $\vect{r}$ of the specified geometry of the particle.
For an arbitrary ensemble of $N$ nanoparticles (particle number indexed with Greek letter $\alpha$) embedded in an uniform dielectric medium, the \textit{collective surface plasmon eigenmodes}, are in turn described with surface charge distributions [denoted with $\Sigma(\vect{r}$)], which can be written as a superposition of the eigenmodes of each particle of the ensemble:
\begin{equation}\label{eq:expansion}
\Sigma(\vect r) = \sum_{\alpha,j} c_{\alpha,j}\,\sigma_{\alpha,j}(\vect r),
\end{equation}
where each coefficient $c_{\alpha,j}$ describes the contribution of the localised surface plasmon resonance $j$ of particle $\alpha$ to the total collective mode of the ensemble.

This collective charge distribution  satisfies an eigenvalue equation \cite{Ouyang_PMPB1989a,Mayergoyz_PRB2005a}:
\begin{equation}\label{eq:surf_eigeneq_N}
\Sigma_n(\vect r) = \frac{\Lambda_n}{2\pi}\oint\,\hat{n}\cdot\frac{(\vect{r}-\vect{r}')}{|\vect{r}-\vect{r}'|^3}\Sigma_n(\vect{r}') \,dS',
\end{equation}
where $\hat{n}$ is a normal vector pointing outwards on a point $\vect{r}$ of the surface $S$ of the nanostructures. 
This equation is simply the surface charge induced by the electric fields from all the nanoparticles, as derived from Coulomb's law.
The eigenvalues $\Lambda_n$ are related to the resonance wavelength (frequency) of the surface plasmon modes of the coupled metal nanoparticles:
\begin{equation}\label{eq:eigenvalue_permittivity}
\Lambda_n = \frac{\epsilon_M(\lambda_n)-\epsilon_b}{\epsilon_M(\lambda_n)+\epsilon_b},    
\end{equation} 
here, $\lambda_n$ is the wavelength of the surface plasmon resonance, $\epsilon_M(\lambda_n)$ is the (wavelength dependent and complex) permittivity of the metal and  $\epsilon_b$ that of the (uniform) background medium.
The relationship is understood to be valid only for the real part of the right--hand side of the equation.

\citet{Gomez_PRB2010a} have previously shown  that the eigenvalue equation \eqref{eq:surf_eigeneq_N} for a system of interacting particles can be written in terms of the non--interacting constituents, a procedure that results in an eigenvalue equation for the coefficients $\vect c$ in equation \eqref{eq:expansion}\cite{Gomez_PRB2010a}:
\begin{equation}\label{eq:coupling_eqn}
\mathbb{K}\cdot\vect c =   \frac{2\pi}{\Lambda}\vect c,
\end{equation}
where the off--diagonal elements of the matrix $\mathbb{K}$ describe the near--field electrostatic interactions between the particles, terms which depend on their separation distance  and  relative orientations.
The diagonal elements are in turn given by the  eigenmodes of the uncoupled particles.
It was also  shown that for geometrically symmetric arrangements of nanoparticles, the collective surface plasmon eigenmodes are simultaneously eigenvectors of equation \eqref{eq:coupling_eqn} and of the symmetry operators that belong to the point group of the interacting ensemble.

For periodic arrangements of metal nanostructures,
it is expected that $\mathbb{K}$ will be invariant with respect to translations by integer multiples $n$ of the lattice constant $\vect{R}$: 
$\mathbb{K}(\vect r) = \mathbb{K}(\vect r + n\vect R)$. 
If we define a translation operator $\mathbb{T}_\vect{R}$, such that its action on a vector $f$ is:
$\mathbb{T}_\vect{R}f(\vect{r}) = f(\vect{r}+\vect{R})$, 
it is then straightforward to show that 
$\mathbb{K}$ and $\mathbb{T}_\vect{R}$
 commute and consequently posses a common set of eigenvectors\cite{Ashcroft_1976a}: these are the collective surface plasmon eigenmodes that we set out to describe in this paper. 
These collective surface plasmon eigenmodes can  be written, in a similar fashion to equation \eqref{eq:expansion}, in terms of particle--centred eigenmodes of the uncoupled members of the periodic structure:
 $\sigma(\vect{r}-\vect{R}_s)$, 
 where $\vect{R}_s$ is the location of the $s$--th lattice point.
Given the periodic nature of the arrangement, a translation by an integer multiple $n$ of the lattice constant $\vect{R}$ leaves  $\sigma(\vect{r}-\vect{R}_s)$ unchanged, namely: 
$\sigma(\vect{r}-\vect{R}_s+n\vect{R}) = \sigma(\vect{r}-\vect{R}_s)$.
The collective eigenmode may however differ by a  phase factor $F$ for a translation by one equivalent point in the lattice, that is: 
$\Sigma(\vect{r}+n\vect{R}) = F^n\Sigma(\vect{r})$.
The exact form of the phase factor depends on the boundary conditions required to describe the periodic structure, as we now proceed to discuss.

\subsection{Finite structures}
\label{sec:EEM_B}

We consider \textit{periodic} structures composed of a \textit{finite} set of $N$ identical unit cells.
Strictly, a finite set of equally--spaced nanostructures is not fully invariant with respect to an arbitrary number of translations. 
However, this class of structures shows many kinds of interesting collective effects.
For this kind of periodic structure, it is reasonable to  expand the collective surface plasmon eigenmodes in terms of unit--cell centred eigenmodes, which by definition, are invariant with respect to a translation by an integer multiple $n$ of the lattice constant $\vect{R}$:
$\sigma(\vect{r}-\vect{R}_s) = \sigma(\vect{r}-n\vect{R}_s)$.
Consequently, a translation operation leads only to  a  \textit{phase shift} $F$ of the collective surface plasmon eigenmode: 
$\Sigma(\vect{r}+n\vect{R}) = F^n\Sigma(\vect{r})$,
with the obvious constraint that $n\in [1,N]$.
Due to the finite extent of the periodic structure, a translation by $N+1$ multiples of the lattice constant is expected to result in:
$\Sigma(\vect{r}+(N+1)\vect{R}) = F^{N+1}\Sigma(\vect{r}) = 0$, 
and this imposes limitations on the  form of the phase factor $F$, which takes on forms akin   to standing--waves.

\subsection{Interaction with light}

Following \citet{Mayergoyz_PRL2007a}, we consider an incident electric field characterised by a wavevector $\vect{k}$   and a frequency $\omega$ given by:
$\vect{E}_n\exp(i\omega t + i\vect{k}\cdot\vect{r}_n)$, 
which on a single nanostructure (located at position $\vect{r}_n$), excites a surface charge distribution 
$\sigma(\vect{r})$ 
that is a superposition of its electrostatic eigenmodes 
$\sigma_n(\vect{r})$: $\sigma(\vect{r}) = \sum_n a_n(\omega)\sigma_n(\vect{r}_n)$, 
where the coefficients 
$a_n(\omega)$ 
of this expansion, termed \textit{the excitation amplitudes}\cite{Davis_ROMP2016a}, are proportional to the dipole moment of the eigenmode 
$\vect{p}_n$, 
and the polarizability per unit volume of the particles 
$f(\omega)$ 
(i.e. $a_n(\omega)\approx f(\omega) \vect{p}_n\cdot \vect{E}_n\exp(i\vect{k}\cdot\vect{r}_n)$).

In the interacting ensemble, the inter--particle interactions modify these excitation amplitudes, and  it can be shown that the resulting excitation amplitudes $\tilde{\vect{a}}$, can be written in terms of those corresponding to the non--interacting  ones $\vect{a}$ according to:
\begin{equation}\label{eq:a}
\tilde{\vect{a}} = \vect{c}^{-T}\cdot(f\vect{\Lambda})^{-1}\vect{c}\cdot \vect{a},
\end{equation}
where $\vect{\Lambda}$ is a (diagonal) matrix of the eigenvalues of equation \eqref{eq:coupling_eqn}.

In the following sections we illustrate the application of the simple analysis presented thus far by finding  solutions to the electrostatic model for cases where the metal nanoparticles are arranged periodically, under a near--neighbour interaction approximation.

\section{Near-neighbour approximation}
\label{sec:NNA}

In the following, for the sake of simplicity (and without loss of generality), we consider only the interaction among single eigenmodes per particle on the periodic structures. 
In reality, the near--field interaction may involve mixing of more than one eigenmode, a case particularly important for structures supporting degenerate eigenmodes.
However, for the case of nanorods of sufficiently large aspect ratios, the single mode approximation is valid because these structures support a strong dipole surface plasmon resonance with a frequency well--separated from those corresponding to higher--order modes.

\subsection{1D chains}
\begin{figure}[tbph!]
\centering
\includegraphics[width=0.9\linewidth]{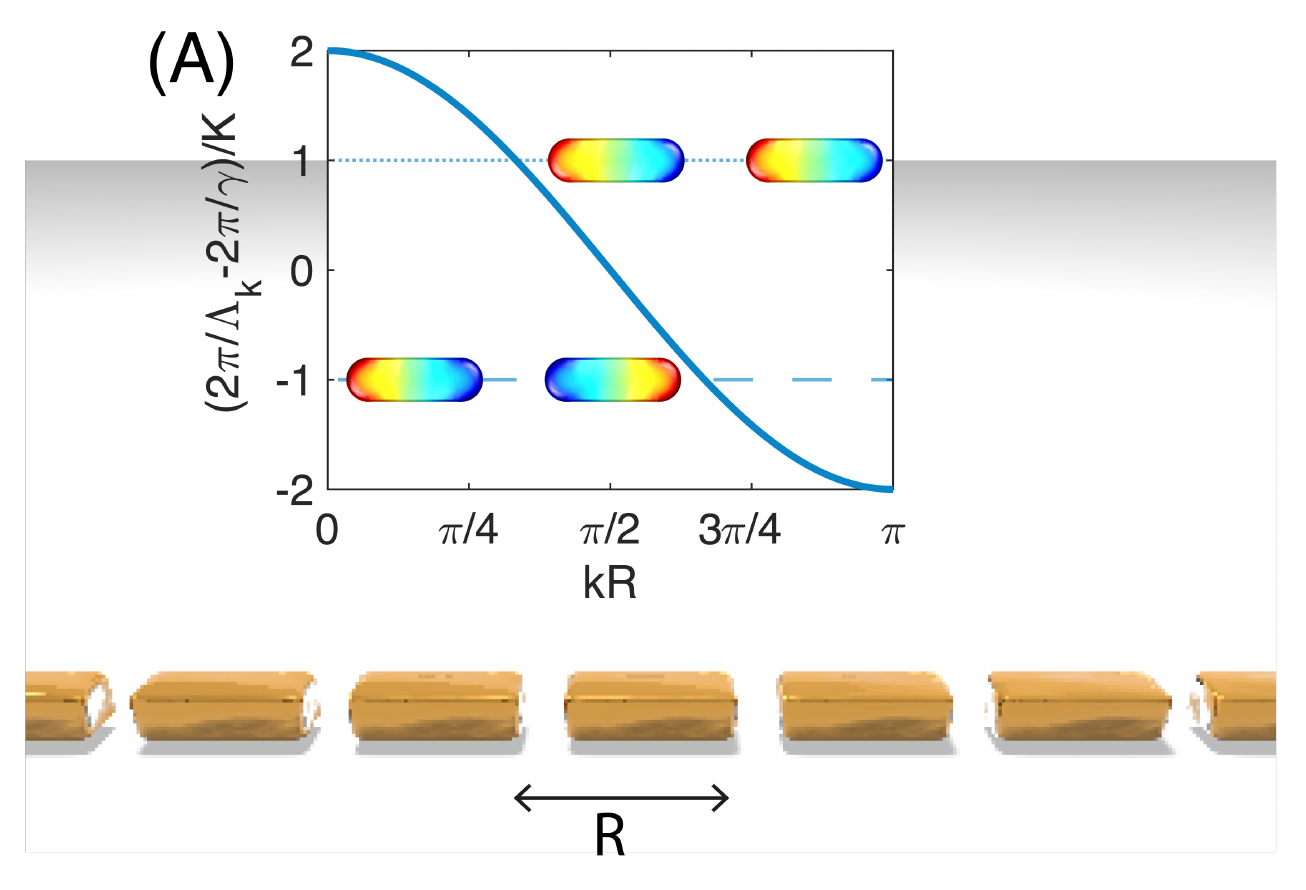}
\caption{Linear chain of nanorods: the diagram shows a set of metal nanorods aligned head to tail with a constant separation distance $R$. 
The inset (A) shows the dispersion relation for the eigenvalues of the chain as given by equation \eqref{eq:1D_finite_eigenvalue}. The dotted and dashed lines correspond to the bright (dotted) and dark (dashed) modes of a dimer, for which the charge distributions are represented. }
\label{fig:1}
\end{figure}

We consider nanoparticle chains such as the one shown in the diagram of Figure 1, which  are characterised by an inter--particle separation distance $R$.
For chains of finite extent,
the collective surface plasmon eigenmode is  written as a superposition of particle--centred eigenmodes: 
$\Sigma(\vect{r}) = \sum_{n}\,c_n\sigma(\vect{r}-\vect{R}_n)$, 
where the sum is carried out for particles $n=1$ to $n=N$ and the lattice points $\vect{R}_n$ are equidistant in one--dimension (i.e. $R_{n+1}-R_n=R$).
With this superposition, and within the nearest--neighbour approximation,
equation \eqref{eq:coupling_eqn} leads to the following recursion relation for the set of coefficients $c_n$ (section S1 shows a next--near--neighbour correction to this approximation):
\begin{equation}
\frac{2\pi}{\Lambda_\vect{k}} c_n = 
c_n\frac{2\pi}{\gamma} + K \left(  c_{n-1}  + c_{n+1}\right),
\end{equation}
where $K$ is the near-neighbour coupling constant, and given the inversion symmetry of the chain, 
$K(\vect R) = K(-\vect R) =K$.  
The finite extent of the chain introduces the requirement that 
$c_0 = c_{N+1} = 0$, 
the  \textit{boundary condition} of the problem

The most general solution to the recursion relation which satisfies the boundary conditions is of the form \cite{Starzak_1989a}:
$c_n = b(\vect{k})\left(e^{i nkR} - e^{-i nkR}\right) = 2 b(\vect{k})\,\sin(n k R)$, directly identifying the phase factor $F$ [$F = 2 \,\sin(n k R)$, $b(\vect{k})$ is a constant that depends on the wavevector $k$ only]discussed in section \ref{sec:EEM_B}.
The boundary conditions of the problem then lead to a discrete set of allowed values for the collective surface plasmon wavevector:
\begin{equation}
k = \frac{m\pi}{(N+1)R},
\end{equation}
with $m=1,2,\dots,N$.
For each wavevector, there is an associated eigenvalue given by:
\begin{equation}\label{eq:1D_finite_eigenvalue}
\frac{2\pi}{\Lambda_k}=\frac{2\pi}{\gamma} + 2K\cos(k R).
\end{equation}
For  large values of $N$, the allowed values of $k$ are closely spaced leading to the eigenvalue dispersion relation  plotted in figure \ref{fig:1}.

For a nanorod dimer, considering only the dipole eigenmode of each rod and for sufficiently large values of $R$, the inter--particle interaction is  dominated by dipole--dipole coupling\cite{Davis_NL2010a} for which it can be shown that  $K>0$.
The electrostatic interaction in a dimer configuration gives rise to two collective eigenmodes whose eigenvalues $\Lambda$ are given by\cite{Davis_NL2010a}
$2\pi/\Lambda = 2\pi/\gamma \pm K$ [represented with dotted horizontal lines in figure \ref{fig:1}(A)], where the \textit{bright mode} is associated with the $+$ sign, corresponds to a situation where the dipoles are pointing in the same direction [see inset of figure \ref{fig:1}(A)] and takes place at wavelengths which are red--shifted from that of the resonance of a single rod [i.e. $\Lambda<\gamma$, top dashed line in figure \ref{fig:1}(A)]. 
The \textit{dark mode} is associated with the $-$ sign, corresponds to the case when the dipoles in the dimer  point in opposite directions and occurs at blue--shifted wavelengths [i.e. $\Lambda>\gamma$, bottom dashed line in figure \ref{fig:1}(A)].
By direct comparison with the result expressed by equation \eqref{eq:1D_finite_eigenvalue}, it can be seen that the effects of 1D periodicity on the eigenvalues are: 
(i) the collective eigenvalues become a  function of the wavevector $k$ of the collective surface plasmon eigenmode, leading to the dispersion curve shown in figure \ref{fig:1}(A), where
the lowest--in--energy collective surface plasmon eigenmode has an associated eigenvector given by 
$2\pi/\Lambda = 2\pi/\gamma + 2K$
(ii) the lowest--in--energy collective surface plasmon eigenmode has an associated wavevector that asymptotes  $k=0$ for increasing values of $N$.
These low--valued $k$ states  physically represent  highly delocalised excitations of the chain where all the dipoles point in one direction: a collective bright mode.
The total dipole moment of this mode is the largest allowed for the 1D chain and consequently, the interaction of the chain with light will be dominated by this delocalised plasmon eigenmode.

According to these results, the lowest energy eigenvalue for an $N$--member chain is given by:
$\frac{1}{\Lambda_N}  =  \frac{1}{\gamma}  + \frac{ K}{\pi}\cos\left(\frac {\pi}{N+1}\right)$, 
and interestingly, this result predicts that in the limit where $N$ approaches infinity,  the resonance frequency (wavelength) of the collective surface plasmon eigenmode reaches an asymptotic value that is only controlled by: 
the geometry of the individual particles (through $\gamma$),   
the magnitude of  the near--neighbour coupling constant $K$,  
the dielectric constant of the uniform surrounding medium.
These predictions have been corroborated experimentally \cite{Barrow_NL2011a}.

\subsubsection{Application to $N=3$}
\begin{figure}[tbph!]
\centering
\includegraphics[width=0.7\linewidth]{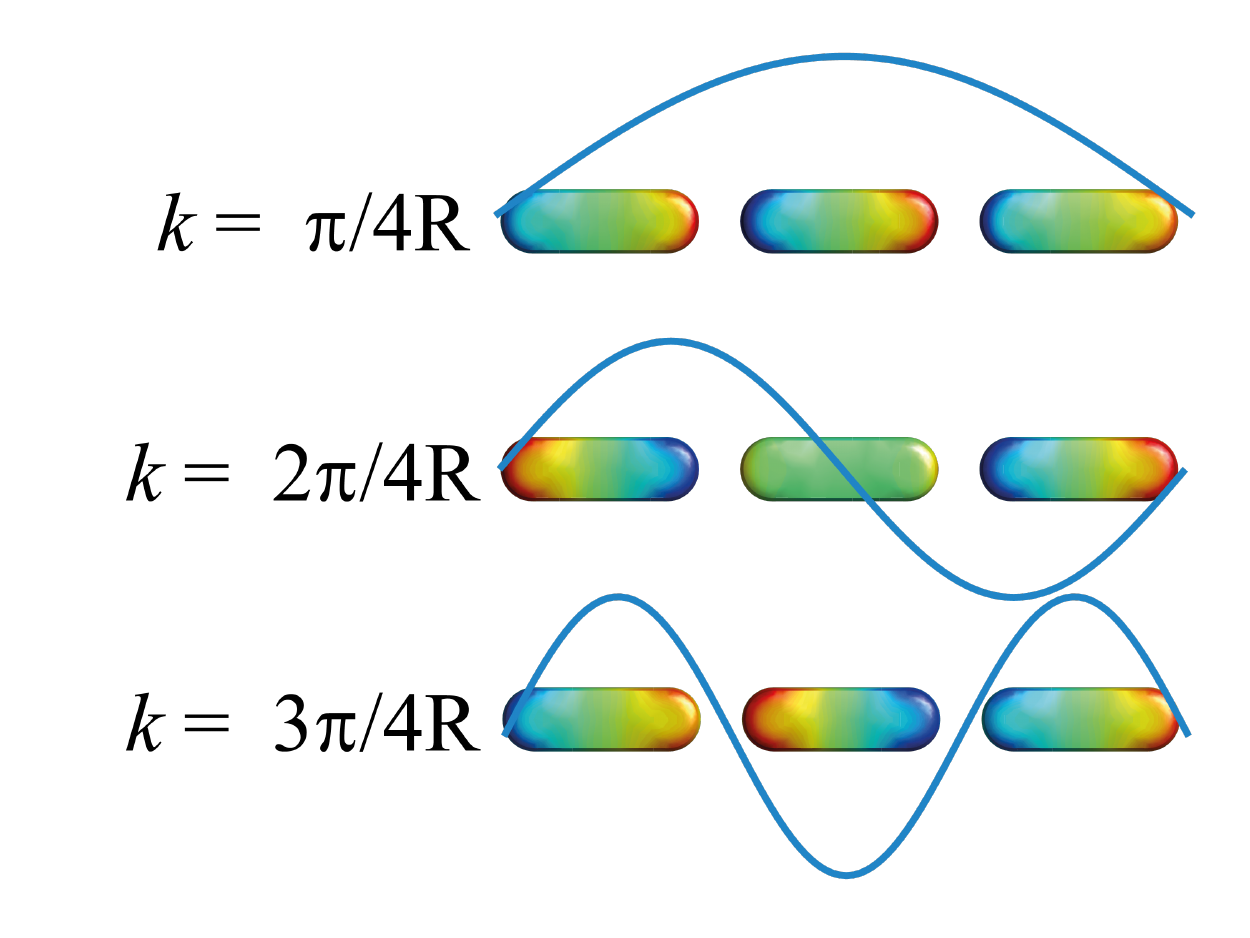}
\caption{Eigenmodes of a chain with $N$=3 particles. The diagram illustrates the collective surface charge distributions (eigenmodes) given by equation \eqref{eq:3mer_eigenmodes}. Each eigenmode has an associated wavevector $k$ which determines the standing--wave--like modulation (shown with solid blue lines) discussed in the text. }
\label{fig:1b}
\end{figure} 
For a simple case consisting of 3 particles, equation \eqref{eq:1D_finite_eigenvalue} gives the following surface plasmon mode eigenvalues and wavevectors:
\begin{equation}
\frac{2\pi}{\Lambda_k}=
\begin{cases}
\frac{2\pi}{\gamma} + 2K/\sqrt{2} & \text{$k=\frac{\pi}{4R}$},\\
\frac{2\pi}{\gamma}               & \text{$k=\frac{2\pi}{4R}$},\\
\frac{2\pi}{\gamma} - 2K/\sqrt{2} & \text{$k=\frac{3\pi}{4R}$},
\end{cases}\nonumber
\end{equation}
with associated collective surface plasmon eigenvectors:
\begin{equation}\label{eq:3mer_eigenmodes}
\Sigma_k(\vect{r}) = 
\begin{cases}
\sigma(\vect{r}-\vect{R}_1) + \sqrt{2}\sigma(\vect{r}-\vect{R}_2) + \sigma(\vect{r}-\vect{R}_3)  & \text{$k=\frac{\pi}{4R}$},\\
\sigma(\vect{r}-\vect{R}_1) - \sigma(\vect{r}-\vect{R}_3)  & \text{$k=\frac{2\pi}{4R}$},\\
\sigma(\vect{r}-\vect{R}_1) - \sqrt{2}\sigma(\vect{r}-\vect{R}_2) + \sigma(\vect{r}-\vect{R}_3)  & \text{$k=\frac{3\pi}{4R}$},
\end{cases}
\end{equation}
which are represented in the diagram of figure \ref{fig:1b}.

The eigenmode associated with  $k=\pi/4R$ corresponds to a situation where all dipoles in the chain  are oriented in the same direction (figure \ref{fig:1b}), resulting in a net dipole moment (bright mode) \cite{Funston_TJOPCL2013b}.
This collective surface plasmon resonance  takes place  at wavelengths red--shifted from that of the uncoupled nanorods ($\Lambda_k<\gamma$), with a redshift which has been shown to increase with the number of particles in the chain \cite{Barrow_NL2011a}. 
Furthermore, the total dipole moment of this collective mode is larger than the simple arithmetic addition of the momenta of each particle (plasmonic super--radiance).

In general, when the chain is made up of an odd number of elements, that is when $N=2j+1$ ($j=1,2,..$), 
there is always a ``zero eigenvalue'' $1/\Gamma_k = 1/\gamma$, which is identical to that of the single particle (more details in section S2).
This eigenvalue has  an associated 
 eigenvector  given by a recurrence relation: $c_{n-1} = -c_{n+1}$, resulting in states of the form $(1,0,-1,0,1,0,-1,..)$, where the notation is meant to represent the coefficients of the  superpositions of particle--centred eigenmodes.
For the trimer of figure \ref{fig:1b}, this state is the one  associated with $k=2\pi/4R$.
These states are important for the discussion of plasmonic edge states that we present later on.


\subsection{Finite chain of dimers: Lattice with a two particle basis} 
\begin{figure}[tbph!]
\centering
\includegraphics[width=0.9\linewidth]{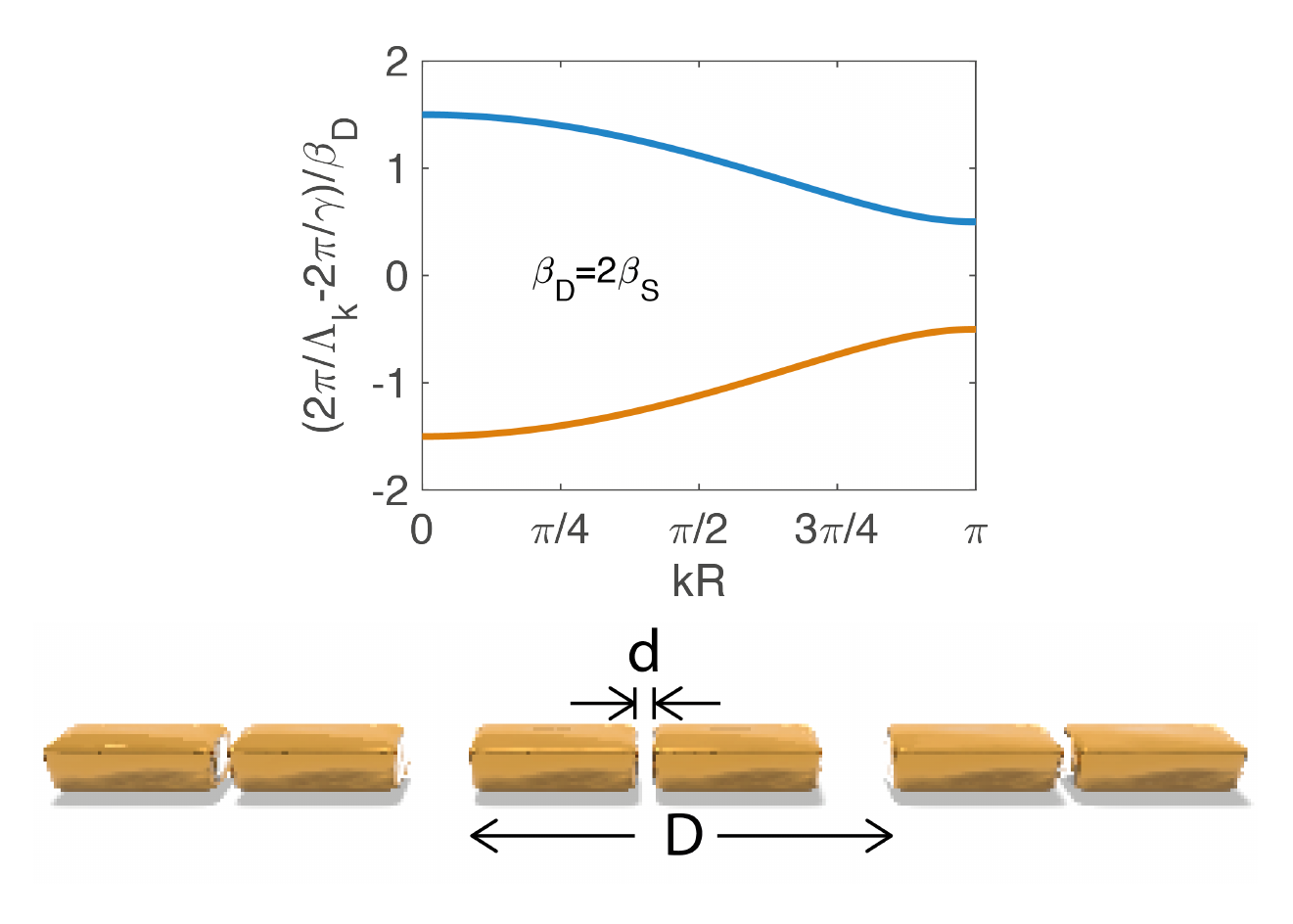}
\caption{Chain of dimers. The diagram shows an infinite chain of dimer characterised by a dimer gap distance of $d$ and a period of $D$. The figure also shows the predicted dispersion curve for the allowed eigenvalues of the collective modes for a particular case where $\beta_D=2\beta_S$.}
\label{fig:2}
\end{figure}

We now consider the case depicted in Figure \ref{fig:2} consisting of a linear chain of particle pairs.
Within each unit cell there is a particle pair with inter--particle separation distance $\vect{d}$.
These particles experience a near--field interaction quantified by a coupling constant $\beta_D$, which is stronger than the interaction $\beta_S$ across adjacent pairs 
(i.e. $\beta_D>\beta_S$) 
due to the fact that the next pair is assumed to be located at a distance $\vect{D}$ with $D>d$.
In this spatial arrangement, the collective surface plasmon eigenmode of the chain is written as  the following superposition of pairs: 
\begin{equation}\label{eq:Sigma_dimer_finite}
\Sigma_k(\vect{r}) = 
\sum_{n=1}^N \left(a_n\sigma(\vect r-\vect{R}_n) + 
b_n\sigma(\vect r-\vect{R}_n-\vect{d})\right),
\end{equation}
where $N$ is the number of pairs in the chain.
Considering only near--neighbour interactions, equation \eqref{eq:surf_eigeneq_N} takes on the following general form:
\begin{equation}\label{eq:2-Toeplitz}
\begin{pmatrix}
0       & \beta_D &0        &\cdots   & 0 \\ 
\beta_D & 0       &\beta_S  &         &   \\
0       & \beta_S &0        & \beta_D &   \\
\vdots  &         & \beta_D & 0       &   \\
0       &         &         &   &\ddots
\end{pmatrix}
\cdot
\begin{pmatrix}
a_1\\b_1\\ \\\vdots\\ b_N
\end{pmatrix}
=
(\frac{2\pi}{\Lambda_k} - \frac{2\pi}{\gamma})
\begin{pmatrix}
a_1\\b_1\\ \\\vdots\\ b_N
\end{pmatrix},
\end{equation}
where $\beta_D$ is the coupling constant between dimers and $\beta_S$ the coupling constant between the closest members of successive  pairs.
These equations correspond to the eigenvalue problem of a tridiagonal 2--Toeplitz matrix \cite{Gover_LAAIA1994a}, for which the eigenvalues and eigenvectors have closed analytical form only for a particular set of conditions\cite{Coulson_POTRSOLAMPAES1938a}.

\citet{Gover_LAAIA1994a} and \citet{Coulson_POTRSOLAMPAES1938a} have demonstrated that for an even number of diagonal elements, the eigenvalues of this problem are  given by:
\begin{equation}\label{eq:2-toeplitz-even}
\frac{2\pi}{\Lambda} = \frac{2\pi}{\gamma} \pm \sqrt{\beta_S^2+\beta_D^2+2\beta_S\beta_D Q_r},
\end{equation}
where $Q_r$ are the zeros of a three point Chebyshev polynomial recurrence formula given in equation (2.31) of \citet{Gover_LAAIA1994a}.

Interestingly, for the  case  when these 2-Toeplitz matrices have an odd number of diagonal elements, \citet{Gover_LAAIA1994a} shows that there is always one eigenvalue given by $1/\Lambda=1/\gamma$.
For these eigenvalues, it is straightforward to show that the coefficients of the eigenvectors, which satisfy the boundary conditions required for the finite extent of the chain, are simply given by:
\begin{equation}\label{eq:edge_state}
\Sigma_0(\vect{r}) \propto \left(1,0,-\frac{\beta_D}{\beta_S},0,\left(-\frac{\beta_D}{\beta_S}\right)^2,0,\dots,0,\left(-\frac{\beta_D}{\beta_S}\right)^n\right),
\end{equation}
which are collective eigenmodes that exhibit localisation towards one end of the chain, a so called \textit{plasmonic edge--state} \cite{Bleckmann_2016a,Wang_NJOP2016a,Ling_OE2015a,Sinev_N2015a,Poddubny_AP2014a,Han_PRL2009a}.

The  remaining $n=N-1$ eigenvalues for an odd--numbered chain of particle pairs are in turn given by:
\begin{equation}\label{eq:2-toeplitz-odd}
\frac{2\pi}{\Lambda} = \frac{2\pi}{\gamma} \pm \sqrt{\beta_S^2+\beta_D^2+2\beta_S\beta_D\cos\left(\frac{2\pi}{n+1}r \right)},
\end{equation}
where $r=1,2,..,(N-1)/2$.

\section{Plasmonic edge states: Application of the formalism to the case of $N$=5}
\label{sec:4}
\begin{figure*}[tbph!]
\centering
\includegraphics[width=0.9\linewidth]{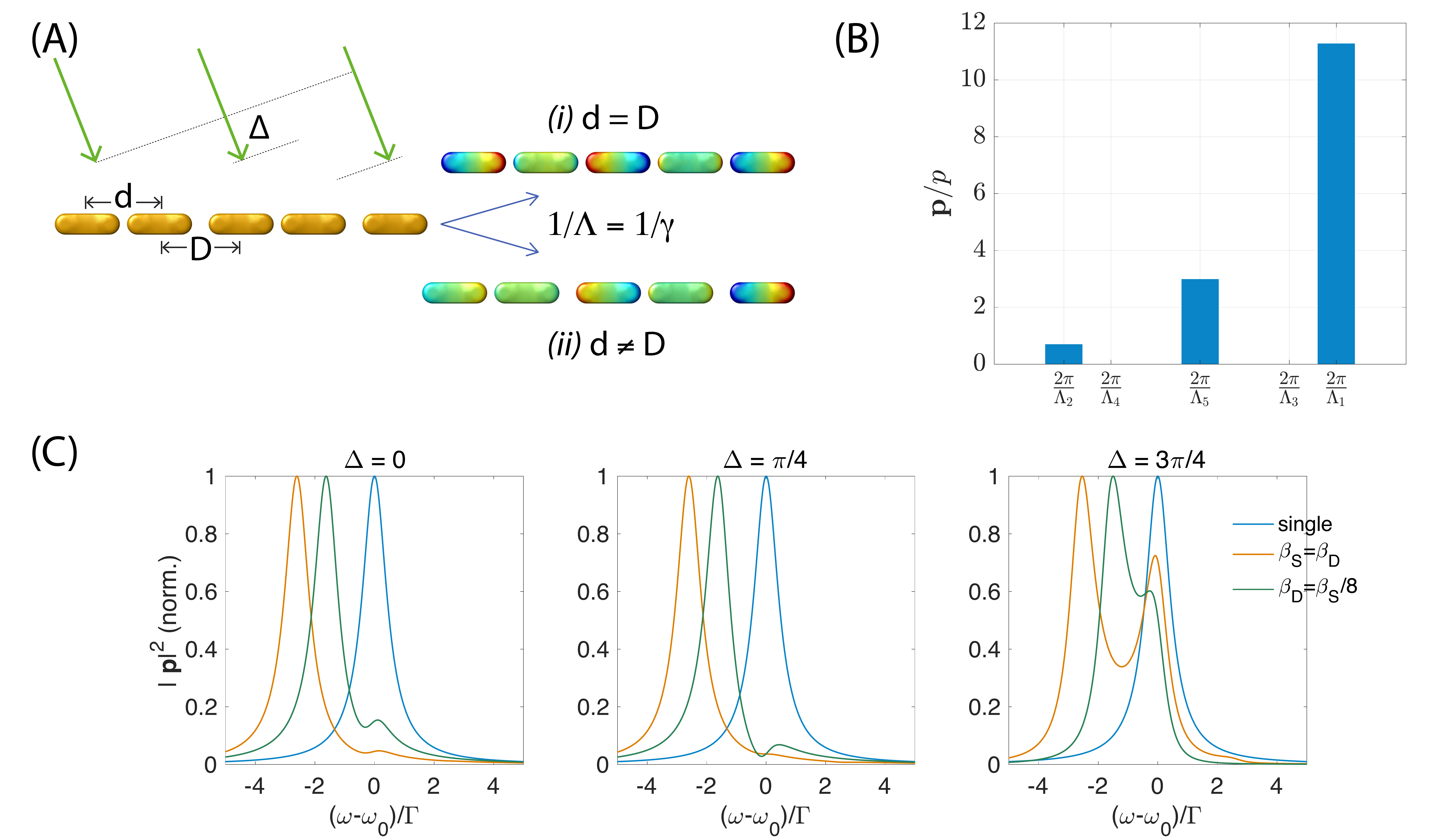}
\caption{Collective plasmonic eigenmodes of nanoparticle chains. 
(A) A chain of 5 nanoparticles is characterised by two inter--particle spacings $d$ and $D$. 
Oblique illumination of this structure (represented with arrows), creates a phase
 difference $\delta$ in the excitation of each member of the chain.  
The surface charge distributions corresponding to the collective eigenmode with eigenvalue  $1/\Lambda = 1/\gamma$ are shown for the cases when: 
(i) $D=d$ and for (ii) $D\ne d$ which exhibits a plasmonic edge state. 
(B) Total dipole moment $\vect{p}$ of the chain for the eigenmodes for the case when $\beta_D=2\beta_S$ expressed in units of $p$ the dipole moment of a single nanorod. 
(C) Calculated squared magnitude of the total dipole moment of a single nanorod and nanorod chains for three phase differences $\Delta$ indicated. }
\label{fig:fig2}
\end{figure*}

We now illustrate the above derivations for a specific example consisting of a chain of 5 identical nanoparticles as shown in figure \ref{fig:fig2}(A), which can be thought of as a chain of dimers with an added particle at one end.
For this configuration, there are five eigenvalues:
four of these are  given by equation \eqref{eq:2-toeplitz-odd}
and,  following our discussion above, the remaining eigenvalue is simply:
$1/\Lambda_5 = 1/\gamma$.
The normalised eigenmode associated with the eigenvalue $\Lambda_5$ is given by:
\begin{equation}\label{eq:vect_loc_5}
\sigma_{\Lambda_5}(\vect r) =
\frac{
\begin{pmatrix}
1,0,-\beta_D/\beta_S, 0, (-\beta_D/\beta_S)^2
\end{pmatrix}}{\sqrt{1+(\beta_S/\beta_D)^2+(\beta_S/\beta_D)^4}}.
\end{equation} 

When $\beta_D>\beta_S$, this collective surface plasmon eigenmode exhibits greater localisation of surface charges on the fifth particle of the  chain at a frequency which corresponds to that of the isolated particle, see figure \ref{fig:fig2}(A) (case ii).
In the figure, it is also shown that when $D=d$ (the simple one--dimensional chain), there is no edge--state, since in this case $\beta_D=\beta_S$, and the analysis follows the one presented in section \ref{sec:EEM_B}.

In figure \ref{fig:fig2}(B) we show the value of the total dipole moment $\vect{p}$ of each of the eigenmodes of the chain 
($\vect{p}=\sum c_np_n = p\sum c_n$).
Two of these eigenmodes have a net zero dipole moment, and interestingly, the plasmonic edge--state (the eigenmode with eigenvalue $\Lambda_5$) has a net dipole moment that is larger than that of a single, non--interacting nanorod.

For this structure, equation \eqref{eq:a} describes its interaction with light.
Following the analysis of \citet{Eftekhari_OL2014a}, and assuming that the optical phase difference $\Delta$ [figure \ref{fig:fig2}(A)] between each dimer is much smaller than the phase difference across dimers, the calculated excitation amplitudes $\tilde{\vect{a}}$ allow for an estimation of the scattering spectrum of the chain, which is in turn proportional to the squared modulus of the total  light--induced dipole moment (see also the supporting information).
This quantity is shown in figure \ref{fig:fig2}(C) as a function of a normalised frequency difference and phase difference ($\Delta$).
Figure \ref{fig:fig2}(C) shows that when the structures are illuminated in phase with each other (i.e. $\Delta=0$, normal incidence), the spectrum for the equally--spaced chains (orange line) is dominated by a spectral band  that is red--shifted from that of the single nanorod [located at $\omega=\omega_0$].
For the chain of dimers with the additional particle (green line), the spectrum also shows a strong band that is also red--shifted with respect to the single nanorod, but it also develops a band at zero detuning [i.e. $\omega=\omega_0$], which corresponds to the excitation of the plasmonic edge eigenmode (the eigenmode with eigenvalue $\Lambda_5$).
The relative contribution of these two bands to the spectrum is sensitive to illumination phase differences as shown in the panels of figure \ref{fig:fig2}(C) (and also in figure S2).

\begin{figure}[tbph!]
\centering
\includegraphics[width=0.9\linewidth]{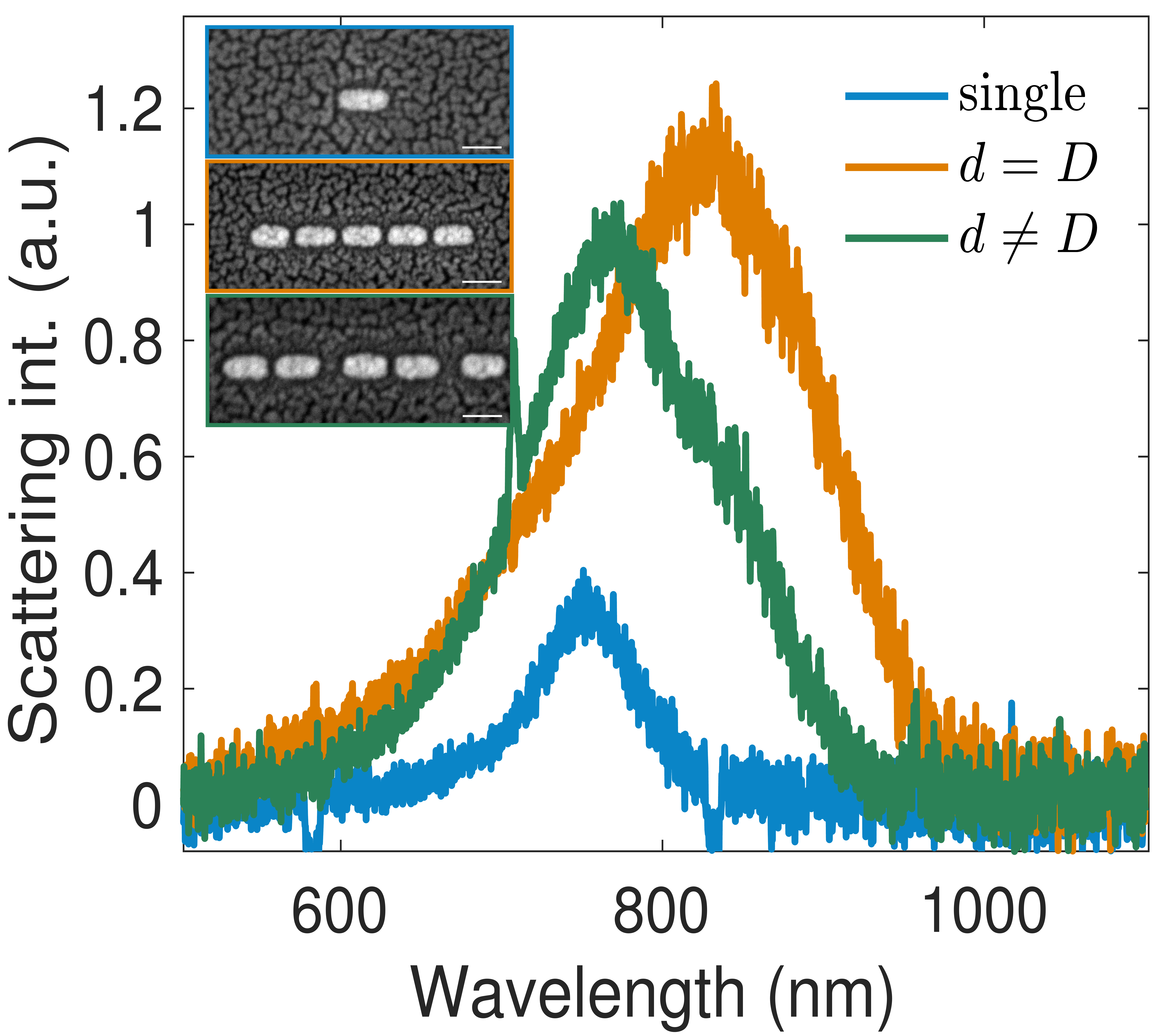}
\caption{Dark--field scattering spectra of: (i) a single Au nanorod of lateral dimensions 100 nm x 40 nm (30 nm height), (ii) a chain of $N$ = 5 nanorods equally spaced ($d=D$), and (iii) a chain with $d\ne D$. The inset shows SEMs of each structure, where the scale bar corresponds to 100 nm. $d \approx $ 20 nm, $D\approx$ 48 nm. In all cases, the substrate was glass.}
\label{fig:edge}
\end{figure}
Figure \ref{fig:edge} shows experimental dark--field scattering spectra measured for  a single Au nanorod, a chain of equally--spaced nanorods and a chain containing two dimers and an additional nanorod. 
In agreement with the predictions of the theory presented in figure \ref{fig:fig2}(C), the scattering spectrum for the equally spaced chain of particles ($d=D$) consists of a single band that is red--shifted and broadened  with respect to that corresponding to a single nanorod (the unit cell).
For the non--equally spaced chain ($d\ne D$), the spectrum has a strong component at the wavelength of the single nanorod.
The experimentally measured spectra were acquired by illuminating the structures with a hollow cone of light with a finite spread of angles of incidence excluding the normal. 
According to the predictions of figure \ref{fig:fig2}(C) (and figure S2), this illumination condition implies that the resulting spectrum will have a strong contribution from the plasmonic edge state (which occurs at the frequency of the single nanorod).
This prediction has been confirmed with finite--difference time domain simulations of the structures of figure \ref{fig:edge} shown in the supporting information section (figure S3).
The FDTD results show: 
(i) strong evidence of the plasmonic edge state through maps of the electric near-field on the chain of nanorods, specifically at the wavelength corresponding to the single nanorod, and 
(ii) the evolution of the calculated scattering spectra agrees well with both the experiment results of figure 5 and is in qualitative agreement with the model predictions of figure 4(C).

Plasmonic edge states in zigzag chains of nanodisks have been both theoretically predicted and experimentally observed \cite{Poddubny_AP2014a,Sinev_N2015a}.
These edge states were found to have a  dependence of the optical response to the number of particles in the chain.
This property arises, in part, due to the mathematical structure of the eigenmode equations (equation \eqref{eq:2-Toeplitz}).
In particular, our theory predicts that these plasmonic edge states occur for situations where: 
(i) $\beta_S \ne \beta_D$, that is when the unit cell of the periodic structure (e.g.  dimer) interacts in a different way to the next near--neighbour, the case taking place in the zigzag configuration which can be viewed as a chain of dimers, and 
(ii) when the number of particles in the chains is odd (leading to the ``zero'' eigenvalue), that is, when a finite chain of dimers ends in a \textit{defect}: a single particle.

\citet{Sinev_N2015a} observed experimentally that  the plasmonic localisation switched between the two ends of the zigzag chains  with changes in the polarization of the incident light.
This observation  can be easily understood with the result shown in equation \eqref{eq:vect_loc_5} (and more generally in equation \eqref{eq:edge_state}): there are two sets of edge eigenstates of the zigzag chains of nanodisks that originate from the two--fold degeneracy of the dipolar eigenmodes of an isolated disk. 
One of the edge eigenstates  stems from the coupling of one kind of  degenerate dipolar eigenmodes (e.g. oriented along axis $\hat x$) and  is characterised by having $\beta_D>\beta_S$: the localisation occurs towards the end of the chain.
The other edge eigenstate arises from the coupling of the orthogonal dipole modes (e.g. those oriented along $\hat y$) leading to $\beta_D<\beta_S$, for which the edge--state localisation occurs in the first element of the chain.
Thus, by changing the polarisation of the incident light it is  possible to control where the localisation takes place along the chain by selective excitation of different (albeit degenerate) edge eigenstates.


\section{Discussion}
In the preceding sections,  we have presented a general electrostatic formulation to describe  the collective surface plasmon eigenmodes of periodic structures.
One of the key elements of the formulation is the \textit{coupling matrix} $\mathbb{K}$ and its off--diagonal elements.
It is important to note that  these off--diagonal elements describe only  near--field interactions which become negligible for separation distances much larger than the wavelength of light.
At larger separation distances, the metal particles can still interact with each other, but in these cases, the interaction is mediated by effects such as far--field radiation and diffraction, which have been shown experimentally to lead to strong modifications to the optical properties of the nanoparticles \cite{Zou_TJOCP2004a,Kravets_PRL2008a, Auguie_PRL2008a,Adato_POTNAOS2009a}.
Our electrostatic formulation does not account for these effects in the description of surface plasmon eigenmodes.

However, the electrostatic formulation can be used for describing these  far--field effects by making explicit consideration of  the interaction of the collective eigenmodes with an incident beam of light of known polarisation.
According to the EEM, the interaction of metallic nanostructures with light is described in terms of electrostatic eigenmodes and their excitation amplitudes\cite{Davis_ROMP2016a,Davis_NL2010a,Davis_PRB2009a} which depend on:
(i) the relative alignment between the incident electric fields and the eigenmode dipole moments,
 and
 (ii) the frequency of the incident field and that of the polarisability of the electrostatic eigenmode, which can be algebraically derived to include the effects of radiative damping. 
 Elements of this approach when applied to periodic structures have been already shown in the work of \citet{Davis_OL2014a}, but a detailed treatment will be presented elsewhere.

In our derivations we have made a deliberate  emphasis on the finite extent of the periodic structures under consideration.
The reason is that the starting point in our derivations, namely equation \eqref{eq:surf_eigeneq_N} is strictly applicable only in the electrostatic limit (size of the nanostructures is smaller than the wavelength of light) and it therefore neglects effects of retardation.
(This is evident in the form of the Green's function that appears under the surface integral).
Previous studies \cite{Davis_OE2009a, Hung_PRB2013a,Davis_ROMP2016a} have shown that main effects of neglecting retardation are underestimation of resonance frequencies and spectral linewidths, 
 for which more appropriate theoretical formalisms, like the Boundary Element method (BEM) \cite{Garcia-de-Abajo_PRB2002a, Hohenester_PRB2005a}, should be considered (it is worth pointing out that equation \eqref{eq:surf_eigeneq_N} is a limiting case of the BEM).
 Although numerically not accurate, the formalism presented here offers physical insight and has a strong predictive power that arises from its simple algebraic structure.

\section{Summary}

In summary, we have presented a theoretical description of the effects of periodicity on the surface plasmon eigenmodes of metal nanostructures.
We have shown applications of the formalism in the description of linear chains of nanostructures. 
In particular, we showed how the method can describe and provide a set of conditions required for the observation of  plasmonic edge states.
We believe that due to the simplicity of this theoretical formulation, it can serve as a basis for designing  complex periodic metal nanostructures for applications in topological photonics, solar energy harvesting, opto--electronics and photocatalysis.

\section{Methods}

\textit{Nanofabrication}
The structures shown in figure \ref{fig:edge} were fabricated by means of electron--beam lithography (Vistec EBPG 5000 plus ES) on glass substrates, using poly(methyl methacrylate) (Micro-Chem, 950k A4) as a resist for a standard lift--off process, and a sacrificial chromium layer (20 nm) that provides for charge dissipation during the exposure.
The structures were developed with a 1:3 (by volume) mixture of  methyl isobutyl ketone/ isopropanol for 90 s, 
rinsed with isopropanol, and dried with a nitrogen gun. 
30 nm of Au were deposited by electron beam evaporation, using 2 nm of chromium as the adhesion layer. 
A subsequent lift off step with acetone produced the nanostructures. 
Normal incidence images of the resulting structures were obtained by scanning electron microscopy (FEI, NovaNanoSEM 430).

\textit{Micro--spectroscopy}
The scattering spectra shown in figure \ref{fig:edge} were acquired with a Nikon Ti-U microscope  coupled to a  Princeton Instrument Isoplane 320  Spectrograph combined with an EMCCD camera.
The light source was a halogen lamp. 
A polarised TU-2 was placed before a dark--field (air) condenser and aligned to the long axis of the structures of figure \ref{fig:edge}, which were imaged using a 40x objective.
The scattering spectra $I_s$ were obtained according to:
\begin{equation}\nonumber
I_s = \frac{I_{NP}-I_B}{I_L - I_B},
\end{equation}
where $I_{NP}$ is the intensity of the scattered light measured from a metal nanostructure, 
$I_B$ the residual background scattering from the substrate and $I_L$ the spectrum of the light source

\section{Acknowledgements}
We acknowledge the ARC for support through a 
Discovery Project (DP160100983)
and a Future Fellowship (FT140100514). 
This work was performed in part at the Melbourne Centre for Nanofabrication (MCN) in the Victorian Node of the Australian National Fabrication Facility (ANFF).

\providecommand{\latin}[1]{#1}
\providecommand*\mcitethebibliography{\thebibliography}
\csname @ifundefined\endcsname{endmcitethebibliography}
  {\let\endmcitethebibliography\endthebibliography}{}

\cleardoublepage

\begin{tocentry} 
\includegraphics[scale=.2]{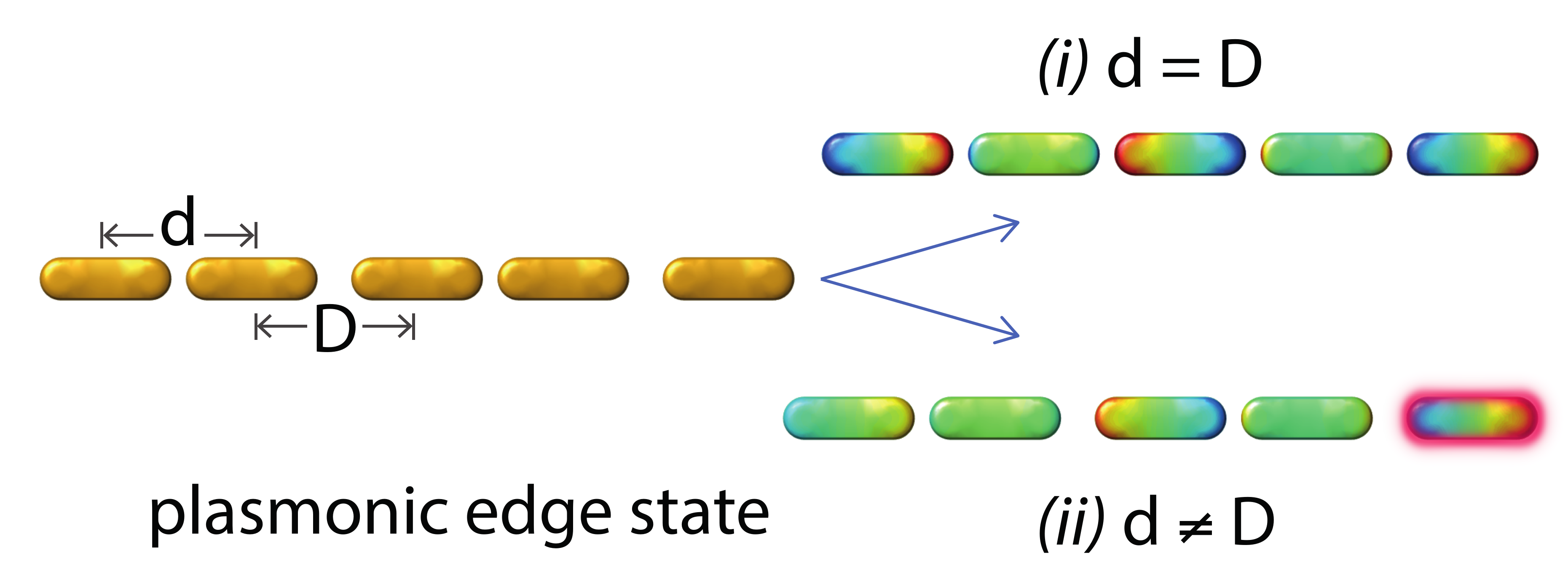} 
\end{tocentry}

\begin{mcitethebibliography}{50}
\providecommand*\natexlab[1]{#1}
\providecommand*\mciteSetBstSublistMode[1]{}
\providecommand*\mciteSetBstMaxWidthForm[2]{}
\providecommand*\mciteBstWouldAddEndPuncttrue
  {\def\EndOfBibitem{\unskip.}}
\providecommand*\mciteBstWouldAddEndPunctfalse
  {\let\EndOfBibitem\relax}
\providecommand*\mciteSetBstMidEndSepPunct[3]{}
\providecommand*\mciteSetBstSublistLabelBeginEnd[3]{}
\providecommand*\EndOfBibitem{}
\mciteSetBstSublistMode{f}
\mciteSetBstMaxWidthForm{subitem}{(\alph{mcitesubitemcount})}
\mciteSetBstSublistLabelBeginEnd
  {\mcitemaxwidthsubitemform\space}
  {\relax}
  {\relax}

\bibitem[Prodan \latin{et~al.}(2003)Prodan, Radloff, Halas, and
  Nordlander]{Prodan_S2003a}
Prodan,~E.; Radloff,~C.; Halas,~N.~J.; Nordlander,~P. {A Hybridization Model
  for the Plasmon Response of Complex Nanostructures}. \emph{Science}
  \textbf{2003}, \emph{302}, 419--422\relax
\mciteBstWouldAddEndPuncttrue
\mciteSetBstMidEndSepPunct{\mcitedefaultmidpunct}
{\mcitedefaultendpunct}{\mcitedefaultseppunct}\relax
\EndOfBibitem
\bibitem[Liu \latin{et~al.}(2009)Liu, Langguth, Weiss, Kastel, Fleischhauer,
  Pfau, and Giessen]{Liu_NM2009a}
Liu,~N.; Langguth,~L.; Weiss,~T.; Kastel,~J.; Fleischhauer,~M.; Pfau,~T.;
  Giessen,~H. Plasmonic analogue of electromagnetically induced transparency at
  the Drude damping limit. \emph{Nat Mater} \textbf{2009}, \emph{8},
  758--762\relax
\mciteBstWouldAddEndPuncttrue
\mciteSetBstMidEndSepPunct{\mcitedefaultmidpunct}
{\mcitedefaultendpunct}{\mcitedefaultseppunct}\relax
\EndOfBibitem
\bibitem[Fano(1961)]{Fano_PR1961a}
Fano,~U. Effects of Configuration Interaction on Intensities and Phase Shifts.
  \emph{Phys. Rev.} \textbf{1961}, \emph{124}, 1866--1878\relax
\mciteBstWouldAddEndPuncttrue
\mciteSetBstMidEndSepPunct{\mcitedefaultmidpunct}
{\mcitedefaultendpunct}{\mcitedefaultseppunct}\relax
\EndOfBibitem
\bibitem[Luk'yanchuk \latin{et~al.}(2010)Luk'yanchuk, Zheludev, Maier, Halas,
  Nordlander, Giessen, and Chong]{Lukyanchuk_NM2010a}
Luk'yanchuk,~B.; Zheludev,~N.~I.; Maier,~S.~A.; Halas,~N.~J.; Nordlander,~P.;
  Giessen,~H.; Chong,~C.~T. The Fano resonance in plasmonic nanostructures and
  metamaterials. \emph{Nat Mater} \textbf{2010}, \emph{9}, 707--715\relax
\mciteBstWouldAddEndPuncttrue
\mciteSetBstMidEndSepPunct{\mcitedefaultmidpunct}
{\mcitedefaultendpunct}{\mcitedefaultseppunct}\relax
\EndOfBibitem
\bibitem[Miroshnichenko \latin{et~al.}(2010)Miroshnichenko, Flach, and
  Kivshar]{Miroshnichenko_RMP2010a}
Miroshnichenko,~A.~E.; Flach,~S.; Kivshar,~Y.~S. Fano resonances in nanoscale
  structures. \emph{Rev. Mod. Phys.} \textbf{2010}, \emph{82}, 2257--2298\relax
\mciteBstWouldAddEndPuncttrue
\mciteSetBstMidEndSepPunct{\mcitedefaultmidpunct}
{\mcitedefaultendpunct}{\mcitedefaultseppunct}\relax
\EndOfBibitem
\bibitem[Guo \latin{et~al.}(2015)Guo, Decker, Setzpfandt, Staude, Neshev, and
  Kivshar]{Guo_NL2015a}
Guo,~R.; Decker,~M.; Setzpfandt,~F.; Staude,~I.; Neshev,~D.~N.; Kivshar,~Y.~S.
  Plasmonic Fano Nanoantennas for On-Chip Separation of Wavelength-Encoded
  Optical Signals. \emph{Nano Letters} \textbf{2015}, \emph{15}, 3324--3328,
  PMID: 25844658\relax
\mciteBstWouldAddEndPuncttrue
\mciteSetBstMidEndSepPunct{\mcitedefaultmidpunct}
{\mcitedefaultendpunct}{\mcitedefaultseppunct}\relax
\EndOfBibitem
\bibitem[Adato \latin{et~al.}(2009)Adato, Yanik, Amsden, Kaplan, Omenetto,
  Hong, Erramilli, and Altug]{Adato_POTNAOS2009a}
Adato,~R.; Yanik,~A.~A.; Amsden,~J.~J.; Kaplan,~D.~L.; Omenetto,~F.~G.;
  Hong,~M.~K.; Erramilli,~S.; Altug,~H. Ultra-sensitive vibrational
  spectroscopy of protein monolayers with plasmonic nanoantenna arrays.
  \emph{Proceedings of the National Academy of Sciences} \textbf{2009},
  \emph{106}, 19227--19232\relax
\mciteBstWouldAddEndPuncttrue
\mciteSetBstMidEndSepPunct{\mcitedefaultmidpunct}
{\mcitedefaultendpunct}{\mcitedefaultseppunct}\relax
\EndOfBibitem
\bibitem[Shelby \latin{et~al.}(2001)Shelby, Smith, and Schultz]{Shelby_S2001a}
Shelby,~R.~A.; Smith,~D.~R.; Schultz,~S. Experimental Verification of a
  Negative Index of Refraction. \emph{Science} \textbf{2001}, \emph{292},
  77--79\relax
\mciteBstWouldAddEndPuncttrue
\mciteSetBstMidEndSepPunct{\mcitedefaultmidpunct}
{\mcitedefaultendpunct}{\mcitedefaultseppunct}\relax
\EndOfBibitem
\bibitem[Schurig \latin{et~al.}(2006)Schurig, Mock, Justice, Cummer, Pendry,
  Starr, and Smith]{Schurig_S2006a}
Schurig,~D.; Mock,~J.~J.; Justice,~B.~J.; Cummer,~S.~A.; Pendry,~J.~B.;
  Starr,~A.~F.; Smith,~D.~R. Metamaterial Electromagnetic Cloak at Microwave
  Frequencies. \emph{Science} \textbf{2006}, \emph{314}, 977--980\relax
\mciteBstWouldAddEndPuncttrue
\mciteSetBstMidEndSepPunct{\mcitedefaultmidpunct}
{\mcitedefaultendpunct}{\mcitedefaultseppunct}\relax
\EndOfBibitem
\bibitem[Ng \latin{et~al.}(2016)Ng, Cadusch, Dligatch, Roberts, Davis,
  Mulvaney, and Gomez]{Ng_AN0a}
Ng,~C.; Cadusch,~J.; Dligatch,~S.; Roberts,~A.; Davis,~T.~J.; Mulvaney,~P.;
  Gomez,~D.~E. Hot Carrier Extraction with Plasmonic Broadband Absorbers.
  \emph{ACS Nano} \textbf{2016}, \emph{10}, 4704--4711\relax
\mciteBstWouldAddEndPuncttrue
\mciteSetBstMidEndSepPunct{\mcitedefaultmidpunct}
{\mcitedefaultendpunct}{\mcitedefaultseppunct}\relax
\EndOfBibitem
\bibitem[Landy \latin{et~al.}(2008)Landy, Sajuyigbe, Mock, Smith, and
  Padilla]{Landy_PRL2008a}
Landy,~N.~I.; Sajuyigbe,~S.; Mock,~J.~J.; Smith,~D.~R.; Padilla,~W.~J. Perfect
  Metamaterial Absorber. \emph{Phys. Rev. Lett.} \textbf{2008}, \emph{100},
  207402\relax
\mciteBstWouldAddEndPuncttrue
\mciteSetBstMidEndSepPunct{\mcitedefaultmidpunct}
{\mcitedefaultendpunct}{\mcitedefaultseppunct}\relax
\EndOfBibitem
\bibitem[Hwang and Davis(2016)Hwang, and Davis]{Hwang_APL2016a}
Hwang,~Y.; Davis,~T.~J. Optical metasurfaces for subwavelength difference
  operations. \emph{Applied Physics Letters} \textbf{2016}, \emph{109}\relax
\mciteBstWouldAddEndPuncttrue
\mciteSetBstMidEndSepPunct{\mcitedefaultmidpunct}
{\mcitedefaultendpunct}{\mcitedefaultseppunct}\relax
\EndOfBibitem
\bibitem[Fang \latin{et~al.}(2005)Fang, Lee, Sun, and Zhang]{Fang_S2005a}
Fang,~N.; Lee,~H.; Sun,~C.; Zhang,~X. Sub--Diffraction--Limited Optical Imaging
  with a Silver Superlens. \emph{Science} \textbf{2005}, \emph{308},
  534--537\relax
\mciteBstWouldAddEndPuncttrue
\mciteSetBstMidEndSepPunct{\mcitedefaultmidpunct}
{\mcitedefaultendpunct}{\mcitedefaultseppunct}\relax
\EndOfBibitem
\bibitem[Hu \latin{et~al.}(2016)Hu, Liu, Ren, Lauhon, and Odom]{Hu_AN2016a}
Hu,~J.; Liu,~C.-H.; Ren,~X.; Lauhon,~L.~J.; Odom,~T.~W. Plasmonic Lattice
  Lenses for Multiwavelength Achromatic Focusing. \emph{ACS Nano}
  \textbf{2016}, \emph{10}, 10275--10282\relax
\mciteBstWouldAddEndPuncttrue
\mciteSetBstMidEndSepPunct{\mcitedefaultmidpunct}
{\mcitedefaultendpunct}{\mcitedefaultseppunct}\relax
\EndOfBibitem
\bibitem[Brongersma \latin{et~al.}(2000)Brongersma, Hartman, and
  Atwater]{Brongersma_PRB2000a}
Brongersma,~M.~L.; Hartman,~J.~W.; Atwater,~H.~A. Electromagnetic energy
  transfer and switching in nanoparticle chain arrays below the diffraction
  limit. \emph{Phys. Rev. B} \textbf{2000}, \emph{62}, R16356--R16359\relax
\mciteBstWouldAddEndPuncttrue
\mciteSetBstMidEndSepPunct{\mcitedefaultmidpunct}
{\mcitedefaultendpunct}{\mcitedefaultseppunct}\relax
\EndOfBibitem
\bibitem[Maier \latin{et~al.}(2003)Maier, Kik, Atwater, Meltzer, Harel, Koel,
  and Requicha]{Maier_NM2003a}
Maier,~S.~A.; Kik,~P.~G.; Atwater,~H.~A.; Meltzer,~S.; Harel,~E.; Koel,~B.~E.;
  Requicha,~A.~A. Local detection of electromagnetic energy transport below the
  diffraction limit in metal nanoparticle plasmon waveguides. \emph{Nat Mater}
  \textbf{2003}, \emph{2}, 229--232\relax
\mciteBstWouldAddEndPuncttrue
\mciteSetBstMidEndSepPunct{\mcitedefaultmidpunct}
{\mcitedefaultendpunct}{\mcitedefaultseppunct}\relax
\EndOfBibitem
\bibitem[Maier \latin{et~al.}(2002)Maier, Kik, and Atwater]{Maier_APL2002a}
Maier,~S.~A.; Kik,~P.~G.; Atwater,~H.~A. Observation of coupled
  plasmon-polariton modes in Au nanoparticle chain waveguides of different
  lengths: Estimation of waveguide loss. \emph{Appl. Phys. Lett.}
  \textbf{2002}, \emph{81}, 1714--1716\relax
\mciteBstWouldAddEndPuncttrue
\mciteSetBstMidEndSepPunct{\mcitedefaultmidpunct}
{\mcitedefaultendpunct}{\mcitedefaultseppunct}\relax
\EndOfBibitem
\bibitem[Maier \latin{et~al.}(2002)Maier, Brongersma, Kik, and
  Atwater]{Maier_PRB2002a}
Maier,~S.~A.; Brongersma,~M.~L.; Kik,~P.~G.; Atwater,~H.~A. Observation of
  near-field coupling in metal nanoparticle chains using far-field polarization
  spectroscopy. \emph{Phys. Rev. B} \textbf{2002}, \emph{65}, 193408\relax
\mciteBstWouldAddEndPuncttrue
\mciteSetBstMidEndSepPunct{\mcitedefaultmidpunct}
{\mcitedefaultendpunct}{\mcitedefaultseppunct}\relax
\EndOfBibitem
\bibitem[Poddubny \latin{et~al.}(2014)Poddubny, Miroshnichenko, Slobozhanyuk,
  and Kivshar]{Poddubny_AP2014a}
Poddubny,~A.; Miroshnichenko,~A.; Slobozhanyuk,~A.; Kivshar,~Y. Topological
  Majorana States in Zigzag Chains of Plasmonic Nanoparticles. \emph{ACS
  Photonics} \textbf{2014}, \emph{1}, 101--105\relax
\mciteBstWouldAddEndPuncttrue
\mciteSetBstMidEndSepPunct{\mcitedefaultmidpunct}
{\mcitedefaultendpunct}{\mcitedefaultseppunct}\relax
\EndOfBibitem
\bibitem[Sinev \latin{et~al.}(2015)Sinev, Mukhin, Slobozhanyuk, Poddubny,
  Miroshnichenko, Samusev, and Kivshar]{Sinev_N2015a}
Sinev,~I.~S.; Mukhin,~I.~S.; Slobozhanyuk,~A.~P.; Poddubny,~A.~N.;
  Miroshnichenko,~A.~E.; Samusev,~A.~K.; Kivshar,~Y.~S. Mapping plasmonic
  topological states at the nanoscale. \emph{Nanoscale} \textbf{2015},
  \emph{7}, 11904--11908\relax
\mciteBstWouldAddEndPuncttrue
\mciteSetBstMidEndSepPunct{\mcitedefaultmidpunct}
{\mcitedefaultendpunct}{\mcitedefaultseppunct}\relax
\EndOfBibitem
\bibitem[Ling \latin{et~al.}(2015)Ling, Xiao, Chan, Yu, and Fung]{Ling_OE2015a}
Ling,~C.~W.; Xiao,~M.; Chan,~C.~T.; Yu,~S.~F.; Fung,~K.~H. Topological edge
  plasmon modes between diatomic chains of plasmonic nanoparticles. \emph{Opt.
  Express} \textbf{2015}, \emph{23}, 2021--2031\relax
\mciteBstWouldAddEndPuncttrue
\mciteSetBstMidEndSepPunct{\mcitedefaultmidpunct}
{\mcitedefaultendpunct}{\mcitedefaultseppunct}\relax
\EndOfBibitem
\bibitem[Wang \latin{et~al.}(2016)Wang, Zhang, Xiao, Han, Chan, and
  Wen]{Wang_NJOP2016a}
Wang,~L.; Zhang,~R.-Y.; Xiao,~M.; Han,~D.; Chan,~C.~T.; Wen,~W. The existence
  of topological edge states in honeycomb plasmonic lattices. \emph{New Journal
  of Physics} \textbf{2016}, \emph{18}, 103029\relax
\mciteBstWouldAddEndPuncttrue
\mciteSetBstMidEndSepPunct{\mcitedefaultmidpunct}
{\mcitedefaultendpunct}{\mcitedefaultseppunct}\relax
\EndOfBibitem
\bibitem[Downing and Weick(2017)Downing, and Weick]{Downing_PRB2017a}
Downing,~C.~A.; Weick,~G. Topological collective plasmons in bipartite chains
  of metallic nanoparticles. \emph{Phys. Rev. B} \textbf{2017}, \emph{95},
  125426\relax
\mciteBstWouldAddEndPuncttrue
\mciteSetBstMidEndSepPunct{\mcitedefaultmidpunct}
{\mcitedefaultendpunct}{\mcitedefaultseppunct}\relax
\EndOfBibitem
\bibitem[Lu \latin{et~al.}(2014)Lu, Joannopoulos, and Soljacic]{Lu_NP2014a}
Lu,~L.; Joannopoulos,~J.~D.; Soljacic,~M. Topological photonics. \emph{Nat
  Photon} \textbf{2014}, \emph{8}, 821--829\relax
\mciteBstWouldAddEndPuncttrue
\mciteSetBstMidEndSepPunct{\mcitedefaultmidpunct}
{\mcitedefaultendpunct}{\mcitedefaultseppunct}\relax
\EndOfBibitem
\bibitem[Davis and G\'omez(2017)Davis, and G\'omez]{Davis_ROMP2016a}
Davis,~T.~J.; G\'omez,~D.~E. \textit{Colloquium:} An algebraic model of
  localized surface plasmons and their interactions. \emph{Reviews of Modern
  Physics} \textbf{2017}, \emph{89}, 011003\relax
\mciteBstWouldAddEndPuncttrue
\mciteSetBstMidEndSepPunct{\mcitedefaultmidpunct}
{\mcitedefaultendpunct}{\mcitedefaultseppunct}\relax
\EndOfBibitem
\bibitem[Gomez \latin{et~al.}(2014)Gomez, Davis, and Funston]{Gomez_JMCC2014a}
Gomez,~D.~E.; Davis,~T.~J.; Funston,~A.~M. Plasmonics by design: design
  principles to structure-function relationships with assemblies of metal
  nanoparticles. \emph{J. Mater. Chem. C} \textbf{2014}, \emph{2},
  3077--3087\relax
\mciteBstWouldAddEndPuncttrue
\mciteSetBstMidEndSepPunct{\mcitedefaultmidpunct}
{\mcitedefaultendpunct}{\mcitedefaultseppunct}\relax
\EndOfBibitem
\bibitem[Davis \latin{et~al.}(2010)Davis, G\'omez, and Vernon]{Davis_NL2010a}
Davis,~T.~J.; G\'omez,~D.~E.; Vernon,~K.~C. Simple Model for the Hybridization
  of Surface Plasmon Resonances in Metallic Nanoparticles. \emph{Nano Letters}
  \textbf{2010}, \emph{10}, 2618--2625\relax
\mciteBstWouldAddEndPuncttrue
\mciteSetBstMidEndSepPunct{\mcitedefaultmidpunct}
{\mcitedefaultendpunct}{\mcitedefaultseppunct}\relax
\EndOfBibitem
\bibitem[Davis \latin{et~al.}(2009)Davis, Vernon, and
  G\'{o}mez]{Davis_PRB2009a}
Davis,~T.~J.; Vernon,~K.~C.; G\'{o}mez,~D.~E. Designing plasmonic systems using
  optical coupling between nanoparticles. \emph{Phys. Rev. B} \textbf{2009},
  \emph{79}, 155423, [also in: Vir. J. Nan. Sci. \& Tech. Vol. 19 (17)
  (2009)]\relax
\mciteBstWouldAddEndPuncttrue
\mciteSetBstMidEndSepPunct{\mcitedefaultmidpunct}
{\mcitedefaultendpunct}{\mcitedefaultseppunct}\relax
\EndOfBibitem
\bibitem[Ashcroft and Mermin(1976)Ashcroft, and Mermin]{Ashcroft_1976a}
Ashcroft,~N.; Mermin,~N. \emph{Solid state physics}; Science: Physics; Saunders
  College, 1976\relax
\mciteBstWouldAddEndPuncttrue
\mciteSetBstMidEndSepPunct{\mcitedefaultmidpunct}
{\mcitedefaultendpunct}{\mcitedefaultseppunct}\relax
\EndOfBibitem
\bibitem[Ouyang and Isaacson(1989)Ouyang, and Isaacson]{Ouyang_PMPB1989a}
Ouyang,~F.; Isaacson,~M. Surface plasmon excitation of objects with arbitrary
  shape and dielectric constant. \emph{Philosophical Magazine Part B}
  \textbf{1989}, \emph{60}, 481--492\relax
\mciteBstWouldAddEndPuncttrue
\mciteSetBstMidEndSepPunct{\mcitedefaultmidpunct}
{\mcitedefaultendpunct}{\mcitedefaultseppunct}\relax
\EndOfBibitem
\bibitem[Mayergoyz \latin{et~al.}(2005)Mayergoyz, Fredkin, and
  Zhang]{Mayergoyz_PRB2005a}
Mayergoyz,~I.~D.; Fredkin,~D.~R.; Zhang,~Z. Electrostatic (plasmon) resonances
  in nanoparticles. \emph{Phys. Rev. B} \textbf{2005}, \emph{72}, 155412\relax
\mciteBstWouldAddEndPuncttrue
\mciteSetBstMidEndSepPunct{\mcitedefaultmidpunct}
{\mcitedefaultendpunct}{\mcitedefaultseppunct}\relax
\EndOfBibitem
\bibitem[G\'omez \latin{et~al.}(2010)G\'omez, Vernon, and
  Davis]{Gomez_PRB2010a}
G\'omez,~D.~E.; Vernon,~K.~C.; Davis,~T.~J. Symmetry effects on the optical
  coupling between plasmonic nanoparticles with applications to hierarchical
  structures. \emph{Phys. Rev. B} \textbf{2010}, \emph{81}, 075414\relax
\mciteBstWouldAddEndPuncttrue
\mciteSetBstMidEndSepPunct{\mcitedefaultmidpunct}
{\mcitedefaultendpunct}{\mcitedefaultseppunct}\relax
\EndOfBibitem
\bibitem[Mayergoyz \latin{et~al.}(2007)Mayergoyz, Zhang, and
  Miano]{Mayergoyz_PRL2007a}
Mayergoyz,~I.~D.; Zhang,~Z.; Miano,~G. Analysis of Dynamics of Excitation and
  Dephasing of Plasmon Resonance Modes in Nanoparticles. \emph{Phys. Rev.
  Lett.} \textbf{2007}, \emph{98}, 147401\relax
\mciteBstWouldAddEndPuncttrue
\mciteSetBstMidEndSepPunct{\mcitedefaultmidpunct}
{\mcitedefaultendpunct}{\mcitedefaultseppunct}\relax
\EndOfBibitem
\bibitem[Starzak(1989)]{Starzak_1989a}
Starzak,~M. \emph{{Mathematical methods in chemistry and physics}}; Plenum Pub
  Corp, 1989\relax
\mciteBstWouldAddEndPuncttrue
\mciteSetBstMidEndSepPunct{\mcitedefaultmidpunct}
{\mcitedefaultendpunct}{\mcitedefaultseppunct}\relax
\EndOfBibitem
\bibitem[Barrow \latin{et~al.}(2011)Barrow, Funston, G\'omez, Davis, and
  Mulvaney]{Barrow_NL2011a}
Barrow,~S.~J.; Funston,~A.~M.; G\'omez,~D.~E.; Davis,~T.~J.; Mulvaney,~P.
  Surface Plasmon Resonances in Strongly Coupled Gold Nanosphere Chains from
  Monomer to Hexamer. \emph{Nano Letters} \textbf{2011}, \emph{11},
  4180--4187\relax
\mciteBstWouldAddEndPuncttrue
\mciteSetBstMidEndSepPunct{\mcitedefaultmidpunct}
{\mcitedefaultendpunct}{\mcitedefaultseppunct}\relax
\EndOfBibitem
\bibitem[Funston \latin{et~al.}(2013)Funston, G{\'o}mez, Karg, Vernon, Davis,
  and Mulvaney]{Funston_TJOPCL2013b}
Funston,~A.~M.; G{\'o}mez,~D.~E.; Karg,~M.; Vernon,~K.~C.; Davis,~T.~J.;
  Mulvaney,~P. Aligned Linear Arrays of Crystalline Nanoparticles. \emph{The
  Journal of Physical Chemistry Letters} \textbf{2013}, \emph{4},
  1994--2001\relax
\mciteBstWouldAddEndPuncttrue
\mciteSetBstMidEndSepPunct{\mcitedefaultmidpunct}
{\mcitedefaultendpunct}{\mcitedefaultseppunct}\relax
\EndOfBibitem
\bibitem[Gover(1994)]{Gover_LAAIA1994a}
Gover,~M. The eigenproblem of a tridiagonal 2-Toeplitz matrix. \emph{Linear
  Algebra and its Applications} \textbf{1994}, \emph{197--198}, 63 -- 78\relax
\mciteBstWouldAddEndPuncttrue
\mciteSetBstMidEndSepPunct{\mcitedefaultmidpunct}
{\mcitedefaultendpunct}{\mcitedefaultseppunct}\relax
\EndOfBibitem
\bibitem[Coulson(1938)]{Coulson_POTRSOLAMPAES1938a}
Coulson,~C.~A. The Electronic Structure of Some Polyenes and Aromatic
  Molecules. IV. The Nature of the Links of Certain Free Radicals.
  \emph{Proceedings of the Royal Society of London A: Mathematical, Physical
  and Engineering Sciences} \textbf{1938}, \emph{164}, 383--396\relax
\mciteBstWouldAddEndPuncttrue
\mciteSetBstMidEndSepPunct{\mcitedefaultmidpunct}
{\mcitedefaultendpunct}{\mcitedefaultseppunct}\relax
\EndOfBibitem
\bibitem[Bleckmann \latin{et~al.}(2016)Bleckmann, Linden, and
  Alberti]{Bleckmann_2016a}
Bleckmann,~F.; Linden,~S.; Alberti,~A. Spectral imaging of topological edge
  states in plasmonic waveguide arrays. 2016\relax
\mciteBstWouldAddEndPuncttrue
\mciteSetBstMidEndSepPunct{\mcitedefaultmidpunct}
{\mcitedefaultendpunct}{\mcitedefaultseppunct}\relax
\EndOfBibitem
\bibitem[Han \latin{et~al.}(2009)Han, Lai, Zi, Zhang, and Chan]{Han_PRL2009a}
Han,~D.; Lai,~Y.; Zi,~J.; Zhang,~Z.-Q.; Chan,~C.~T. Dirac Spectra and Edge
  States in Honeycomb Plasmonic Lattices. \emph{Phys. Rev. Lett.}
  \textbf{2009}, \emph{102}, 123904\relax
\mciteBstWouldAddEndPuncttrue
\mciteSetBstMidEndSepPunct{\mcitedefaultmidpunct}
{\mcitedefaultendpunct}{\mcitedefaultseppunct}\relax
\EndOfBibitem
\bibitem[Eftekhari \latin{et~al.}(2014)Eftekhari, G\'{o}mez, and
  Davis]{Eftekhari_OL2014a}
Eftekhari,~F.; G\'{o}mez,~D.~E.; Davis,~T.~J. Measuring subwavelength phase
  differences with a plasmonic circuit: an example of nanoscale optical signal
  processing. \emph{Opt. Lett.} \textbf{2014}, \emph{39}, 2994--2997\relax
\mciteBstWouldAddEndPuncttrue
\mciteSetBstMidEndSepPunct{\mcitedefaultmidpunct}
{\mcitedefaultendpunct}{\mcitedefaultseppunct}\relax
\EndOfBibitem
\bibitem[Zou and Schatz(2004)Zou, and Schatz]{Zou_TJOCP2004a}
Zou,~S.; Schatz,~G.~C. Narrow plasmonic/photonic extinction and scattering line
  shapes for one and two dimensional silver nanoparticle arrays. \emph{The
  Journal of Chemical Physics} \textbf{2004}, \emph{121}, 12606--12612\relax
\mciteBstWouldAddEndPuncttrue
\mciteSetBstMidEndSepPunct{\mcitedefaultmidpunct}
{\mcitedefaultendpunct}{\mcitedefaultseppunct}\relax
\EndOfBibitem
\bibitem[Kravets \latin{et~al.}(2008)Kravets, Schedin, and
  Grigorenko]{Kravets_PRL2008a}
Kravets,~V.~G.; Schedin,~F.; Grigorenko,~A.~N. Extremely Narrow Plasmon
  Resonances Based on Diffraction Coupling of Localized Plasmons in Arrays of
  Metallic Nanoparticles. \emph{Phys Rev Lett} \textbf{2008}, \emph{101},
  087403\relax
\mciteBstWouldAddEndPuncttrue
\mciteSetBstMidEndSepPunct{\mcitedefaultmidpunct}
{\mcitedefaultendpunct}{\mcitedefaultseppunct}\relax
\EndOfBibitem
\bibitem[Augui\'e and Barnes(2008)Augui\'e, and Barnes]{Auguie_PRL2008a}
Augui\'e,~B.; Barnes,~W.~L. Collective Resonances in Gold Nanoparticle Arrays.
  \emph{Phys. Rev. Lett.} \textbf{2008}, \emph{101}, 143902\relax
\mciteBstWouldAddEndPuncttrue
\mciteSetBstMidEndSepPunct{\mcitedefaultmidpunct}
{\mcitedefaultendpunct}{\mcitedefaultseppunct}\relax
\EndOfBibitem
\bibitem[Davis \latin{et~al.}(2014)Davis, G\'{o}mez, and
  Eftekhari]{Davis_OL2014a}
Davis,~T.~J.; G\'{o}mez,~D.~E.; Eftekhari,~F. All-optical modulation and
  switching by a metamaterial of plasmonic circuits. \emph{Opt. Lett.}
  \textbf{2014}, \emph{39}, 4938--4941\relax
\mciteBstWouldAddEndPuncttrue
\mciteSetBstMidEndSepPunct{\mcitedefaultmidpunct}
{\mcitedefaultendpunct}{\mcitedefaultseppunct}\relax
\EndOfBibitem
\bibitem[Davis \latin{et~al.}(2009)Davis, Vernon, and G\'{o}mez]{Davis_OE2009a}
Davis,~T.~J.; Vernon,~K.~C.; G\'{o}mez,~D.~E. Effect of retardation on
  localized surface plasmon resonances in a metallic nanorod. \emph{Opt.
  Express} \textbf{2009}, \emph{17}, 23655--23663\relax
\mciteBstWouldAddEndPuncttrue
\mciteSetBstMidEndSepPunct{\mcitedefaultmidpunct}
{\mcitedefaultendpunct}{\mcitedefaultseppunct}\relax
\EndOfBibitem
\bibitem[Hung \latin{et~al.}(2013)Hung, Lee, McGovern, Rabin, and
  Mayergoyz]{Hung_PRB2013a}
Hung,~L.; Lee,~S.~Y.; McGovern,~O.; Rabin,~O.; Mayergoyz,~I. Calculation and
  measurement of radiation corrections for plasmon resonances in nanoparticles.
  \emph{Phys. Rev. B} \textbf{2013}, \emph{88}, 075424\relax
\mciteBstWouldAddEndPuncttrue
\mciteSetBstMidEndSepPunct{\mcitedefaultmidpunct}
{\mcitedefaultendpunct}{\mcitedefaultseppunct}\relax
\EndOfBibitem
\bibitem[Garc\'{\i}a~de Abajo and Howie(2002)Garc\'{\i}a~de Abajo, and
  Howie]{Garcia-de-Abajo_PRB2002a}
Garc\'{\i}a~de Abajo,~F.~J.; Howie,~A. Retarded field calculation of electron
  energy loss in inhomogeneous dielectrics. \emph{Phys Rev B} \textbf{2002},
  \emph{65}\relax
\mciteBstWouldAddEndPuncttrue
\mciteSetBstMidEndSepPunct{\mcitedefaultmidpunct}
{\mcitedefaultendpunct}{\mcitedefaultseppunct}\relax
\EndOfBibitem
\bibitem[Hohenester and Krenn(2005)Hohenester, and Krenn]{Hohenester_PRB2005a}
Hohenester,~U.; Krenn,~J. Surface plasmon resonances of single and coupled
  metallic nanoparticles: A boundary integral method approach. \emph{Phys. Rev.
  B} \textbf{2005}, \emph{72}, 195429\relax
\mciteBstWouldAddEndPuncttrue
\mciteSetBstMidEndSepPunct{\mcitedefaultmidpunct}
{\mcitedefaultendpunct}{\mcitedefaultseppunct}\relax
\EndOfBibitem
\end{mcitethebibliography}
\end{document}